\documentclass[manuscript]{acmart}

\AtBeginDocument{%
  }

\copyrightyear{2022} 
\acmYear{2022} 
\setcopyright{rightsretained} 
\acmConference[RecSys '22]{Sixteenth ACM Conference on Recommender Systems}{September 18--23, 2022}{Seattle, WA, USA}
\acmBooktitle{Sixteenth ACM Conference on Recommender Systems (RecSys '22), September 18--23, 2022, Seattle, WA, USA}
\acmDOI{10.1145/3523227.3547386}
\acmISBN{978-1-4503-9278-5/22/09}

\usepackage{subcaption}
\begin{document}

\title{Taxonomic Recommendations of Real Estate Properties with Textual Attribute Information}

\author{Zachary Harrison}
\authornote{Both authors contributed equally to this research.}
\email{zacharyha@zillowgroup.com}
\author{Anish Khazane}
\authornotemark[1]
\email{anishk@zillowgroup.com}
\affiliation{%
  \institution{Zillow Group}
  \streetaddress{1301 Second Avenue Floor 31}
  \city{Seattle}
  \state{WA}
  \country{USA}
  \postcode{98101}
}

\begin{abstract}
  In this extended abstract, we present an end to end approach for building a taxonomy of home attribute terms that enables hierarchical recommendations of real estate properties. We cover the methodology for building a real-estate taxonomy, metrics for measuring this structure's quality, and then conclude with a production use-case of making recommendations from search keywords at different levels of topical similarity.
\end{abstract}



\maketitle

\enlargethispage{-12pt}

\section{Introduction and Motivation}
One of Zillow's primary objectives is to provide customers with easily accessible and relevant information for millions of listings on our platform. Our recently launched home insights feature enables this vision by highlighting the most unique keyphrases from a property's description on their respective home details page.

These tags can cover a wide array of attribute types, such as interior (e.g hardwood floors), exterior (e.g outdoor patio) or even community or location attributes (e.g nearby golf course) that can be useful for users searching for specific features in homes. Categorizing properties simply based on these raw phrases is difficult due to the significant variety in possible topics. However, organizing these terms in a taxonomy structure can not only help with categorizing these terms under different levels of topical similarity but also enable hierarchical recommendations. 

In this extended abstract, we present an approach for constructing a real-estate specific taxonomy of home attribute terms that can enable hierarchical recommendations of property listings. We first describe a semi-automated approach to construct this taxonomy, propose metrics for measuring edge quality, then conclude with a production use-case of making hierarchical recommendations from search keywords at different levels of topical similarity. We hope this work can contribute to the few examples of tree-based recommender systems used in industry such as Pinterest's pin2interest knowledge graph system and Yahoo's taxonomy powered recommender systems for modeling user purchase behavior \cite{yahoorecommender, pin2interest}.


\section{Methodology and Challenges}
\begin{table*}
  \caption{State of taxonomy after each round of expansion detailed in Section 2.1-2.2. Link pruning and human-in-the-loop revision is critical for constructing a deeper taxonomy structure for hierarchical recommendations.}
  \label{tab:freq}
  \begin{tabular}{cccccc}
    \toprule
    \multicolumn{1}{p{2cm}}{\centering \ Expansion Stage} &
    \multicolumn{1}{p{2cm}}{\centering \# Nodes} & \multicolumn{1}{p{2cm}}{\centering \# Edges} & \multicolumn{1}{p{2cm}}{\centering \# Parents} & \multicolumn{1}{p{2cm}}{\centering \# Leaf Nodes} & \multicolumn{1}{p{2cm}}{\centering Max Depth}\\
    \midrule
    \multicolumn{1}{p{2cm}}{\centering Seed taxonomy} &
    \multicolumn{1}{p{2cm}}{\centering 2560} & \multicolumn{1}{p{2cm}}{\centering 2559} & \multicolumn{1}{p{2cm}}{\centering 50}
    & \multicolumn{1}{p{2cm}}{\centering 2509} & \multicolumn{1}{p{2cm}}{\centering 2} \\
    \midrule
    \multicolumn{1}{p{2cm}}{\centering Embedding Clustering} &
    \multicolumn{1}{p{2cm}}{\centering 4023} & \multicolumn{1}{p{2cm}}{\centering 4022} & \multicolumn{1}{p{2cm}}{\centering 50}
    & \multicolumn{1}{p{2cm}}{\centering 3972} & \multicolumn{1}{p{2cm}}{\centering 2} \\
     \midrule
    \multicolumn{1}{p{2cm}}{\centering Link Pruning and Manual Revision} &
    \multicolumn{1}{p{2cm}}{\centering 9138} & \multicolumn{1}{p{2cm}}{\centering 9137} & \multicolumn{1}{p{2cm}}{\centering 806}
    & \multicolumn{1}{p{2cm}}{\centering 8332} & \multicolumn{1}{p{2cm}}{\centering 7} \\
  \bottomrule
\end{tabular}
\label{taxonomydetails}
\Description{This table details the structure of the taxonomy after each round of ontology expansion detailed in Sections 2.1-2.2.}
\end{table*}

\subsection{Bootstrapping a Seed Taxonomy}
\label{seedtaxonomy}

We begin by using a pretrained scene entity detection network to label thousands of photos spanning many types of properties (e.g single family homes, apartment rentals, condos) with high-level concept attributes like “KITCHEN” or “BATHROOM”, and also a generative model to tag images with keywords (e.g “granite countertops”, “wood flooring”) describing the contents of the image \cite{zillowtrulia}. Thus, we can rapidly bootstrap a 2-level taxonomy by mapping the keywords of an image to the high-level scene entities identified by the deep network. This approach yields 2560 nodes and 2559 edges as displayed in Table \ref{taxonomydetails}.  


\enlargethispage{-12pt}


We then expand this structure by training a fasttext embedding model on listing descriptions to generate subword embedding representations of the aforementioned home insight phrases and scene entities in the taxonomy \cite{fasttext}. Following embedding generation, we initialize a k-nearest neighbors model with the scene entity embeddings and then classify each home insight embedding if the closest neighbor (k=1)  exceeds a cosine similarity threshold $\alpha$. We use $\alpha$ equal to 0.80 after optimizing for the precision metric defined in Table \ref{precisiontable}.

As seen in Table \ref{taxonomydetails}, the embedding clustering approach yields a significant increase in the total number nodes and edges in the taxonomy. However, there will still be many noisy parent to child relationships due to the disadvantage of using only textual similarity for link construction. For example, a keyphrase  “golf” may become a child under the category “golf course” because the cosine similarity between these two phrase embeddings is above the $\alpha$ threshold, but this would not be a useful link for categorization purposes. Thus, we use the taxonomy at this stage to bootstrap training data for the link pruning algorithm described in the following section, which will aim to mitigate some of these improperly defined edges in the taxonomy.

\subsection{Link Pruning and Revision}
\label{linkpruning}
In order to remove noisy edges in the taxonomy, we set up a binary classification task to predict the probability of a valid parent-child relationship. Formally, we define a node $i$, its parent as $p(i)$, and an edge connecting the two nodes as $e(i,p(i))$ In addition, a candidate parent is defined as a node in the taxonomy with at least one child node.

We first manually review all edges from the taxonomy generated from Section 2.1 to remove any clearly incorrect parent-child relationships. We then use the remaining edges as positive samples for the binary classification task, yielding roughly 4000 pairs represented as $(token_{i} [SEP] token_{p(i)}, 1)$ The inputs $token_{i}$ and $token_{p(i)}$ are the tokenized node tags separated by the special token $[SEP]$. $1$ corresponds to the positive output label. We then create a similar number of negative samples by creating tuples $(token_{i} [SEP] token_{p(j)}, 0)$ where $p(j)$ is a candidate parent that does not lie on the path from a node $i$ to the root node of the taxonomy. 
 
Following data generation, we train a BERT-based binary text classification model on these tuples as input and remove all child nodes $i$ from edges $e(i, p(i))$ which are classified as invalid \cite{bert}. For each pruned child node $k$, we create pairs with all possible candidate parents in the taxonomy and then apply the aforementioned classification model to get a probability representing the validity of each possible relationship. We then create an edge between the node $k$ and the candidate with the maximum probability of being a parent. The last row of Table \ref{taxonomydetails} displays the final state of the taxonomy after this approach.

We can use this inference protocol to automatically suggest additional edges in the taxonomy if there are new home insights on listings, which can then be manually approved or rejected by a reviewer. This human-in-the-loop step is critical for maintaining highly precise edges for the candidate recommendations described in Section \ref{recommendationssection} and a common feature of other taxonomy-based systems in literature \cite{pin2interest}.

\begin{table}
  \caption{WordNet Precision}
  \begin{tabular}{cc}
    \toprule
    \multicolumn{1}{p{3cm}}{\centering Edge Set} & \multicolumn{1}{p{2cm}}{\centering Precision}\\
    \midrule
    \multicolumn{1}{p{3cm}}{\centering Random} & \multicolumn{1}{p{2cm}}{\centering 0.0}\\
    \midrule
    \multicolumn{1}{p{3cm}}{\centering Embedding Similarity} & \multicolumn{1}{p{2cm}}{\centering 0.114}\\
    \midrule
    \multicolumn{1}{p{3cm}}{\centering Taxonomy} & \multicolumn{1}{p{2cm}}{\centering 0.211}\\
  \bottomrule
\end{tabular}
\label{precisiontable}
\Description{This table evaluates the quality of edges in the taxonomy by comparing the percentage of its edges that appear in WordNet with the percentage of edges generated via random sampling or embedding similarity.}
\end{table}

\section{METRICS}
\subsection{Quality of Taxonomy Relationships}

In order to validate the quality of edges in the taxonomy, we look at all nodes in the structure that also appear in WordNet, which is a publicly available ontology linking words into semantic relations with different constructs (e.g synonyms). For each node in this subset, we evaluate three distinct approaches for edge construction (i) creating an edge between the node and a random candidate parent from the taxonomy (ii) creating an edge between the node and the candidate parent with closest embedding similarity and (iii) using the existing relationship defined in the taxonomy. 

We then calculate precision by computing the number of edges from each of the approaches (i-iii) that exist in WordNet over the total number of possible WordNet edges that include a node from the taxonomy. As seen in Table \ref{precisiontable}, there is a clear improvement from building a hierarchical structure to capture more complicated WordNet relationships like hyponyms and meronyms.

\subsection{Evaluation of taxonomy-based Candidate Selection for Recommendations}
\label{recommendationssection}

Table \ref{recommendations} compares a baseline substring match algorithm versus the taxonomy for generating candidate recommendations for top search keywords in the Seattle metropolitan area. We focus on this region due to the abundance of listings (20,000+) that are available for candidate selection.

\begin{table*}
  \caption{Number of Candidate Recommendations for Top Search Keywords in Seattle}
  \begin{tabular}{cccc}
    \toprule    
    \multicolumn{1}{p{2cm}}{\centering Popular Search \\ Keyword} & \multicolumn{1}{p{2cm}}{\centering Baseline \\ (substring match)} & \multicolumn{1}{p{2cm}}{\centering taxonomy \\ (Parent)}
    & \multicolumn{1}{p{2cm}}{\centering taxonomy \\ (Grandparent)}\\
    \midrule
    gym & 162 &  \multicolumn{1}{p{2cm}}{\centering 2222 \\ (Gyms and Studio)} & \multicolumn{1}{p{2cm}}{\centering 3811 \\ (Sports and Recreation) } \\
    \hline 
    mid century & 31 &  \multicolumn{1}{p{2cm}}{\centering 98 \\ (Vintage)} & \multicolumn{1}{p{2cm}}{\centering 8702 \\ (Style) } \\
    \hline
    houseboat & 2 &  \multicolumn{1}{p{2cm}}{\centering 103 \\ (Boat Dock)} & \multicolumn{1}{p{2cm}}{\centering 5806 \\ (Waterfront) } \\
    \hline
    balcony & 536 &  \multicolumn{1}{p{2cm}}{\centering 733 \\ (Loft)} & \multicolumn{1}{p{2cm}}{\centering 1777 \\ (Rooms) } \\ 
  \bottomrule
\end{tabular}
\label{recommendations}
\Description{This table compares the number of candidate recommendations that can be generated via the taxonomy with a baseline substring match approach.}
\end{table*}

On each property listing, “home insights” are keyphrases under descriptions that highlight interesting attribute information about the property (e.g “swimming pool”, “wood flooring”). The substring match algorithm counts a listing as a potential candidate if any substring of home insights associated with the listing matches with the search keyword in question. Column 2 of Table \ref{recommendations} displays the total number of candidate recommendations from using this approach. While this method finds a significant number of candidates for simple queries (e.g gym, balcony), it is unsurprisingly not effective for generating candidates for more complicated query terms (e.g mid century, houseboat). 

Columns 3 and 4 display the total number of candidate listings from using two resolutions of categorization with the taxonomy, with the following algorithm (i)
map each home insight tag on a listing to the closest category in the taxonomy, (ii) 
map the search keyphrase to the closest category in the taxonomy, (iii) 
count a listing as a candidate if there is any intersection of categories between the output from 1 and 2. 

Unsurprisingly, we find a significant increase in the total number of candidate listings over the baseline approach for all top search terms in Table \ref{recommendations}. The more interesting takeaway from this comparison is the ability to produce candidates at different resolutions of similarity with an taxonomy-based structure. For example, “gym” could be mapped to a cluster of candidate listings with home insights categorized under “Gyms and Studios” or could be mapped to a more general cluster of listings categorized under “Sports and Recreation.” 

The ability to generate candidates that may only be loosely related to a search query (in terms of substring matches or direct embedding similarity) but still fall under a similar topic of interest is extremely valuable for diversifying recommendations. This is particularly crucial for a real-estate application where customers are typically unsure about what supply is available to them. Furthermore, unlike more traditional topic modeling approaches like Latent Dirichlet allocation, an taxonomy-based approach also provides a hierarchical ordering of topics that can also be dynamically adjusted based on the requirements of the recommender system.

\section{Conclusion and Future Work}

In this extended abstract, we present an end-to-end approach for constructing a taxonomy for hierarchical recommendations of listing properties. We discuss how a hierarchical structure can capture more sophisticated edge relationships (e.g hyponyms) compared to simple embedding similarity as well as how a taxonomy can increase the total number of candidate recommendations for a real-estate search application. There are several areas of future work including (ii) exploring active learning techniques to propose new edges in the taxonomy based on recently added keywords to listings, and (ii) porting the taxonomy into an ontology-based structure that could enable recommendations across child-child, child-entity, and other relationships not explored in this work.




\bibliographystyle{ACM-Reference-Format}
\bibliography{paper}


\begin{thebibliography}{5}


\ifx \showCODEN    \undefined \def \showCODEN     #1{\unskip}     \fi
\ifx \showDOI      \undefined \def \showDOI       #1{#1}\fi
\ifx \showISBNx    \undefined \def \showISBNx     #1{\unskip}     \fi
\ifx \showISBNxiii \undefined \def \showISBNxiii  #1{\unskip}     \fi
\ifx \showISSN     \undefined \def \showISSN      #1{\unskip}     \fi
\ifx \showLCCN     \undefined \def \showLCCN      #1{\unskip}     \fi
\ifx \shownote     \undefined \def \shownote      #1{#1}          \fi
\ifx \showarticletitle \undefined \def \showarticletitle #1{#1}   \fi
\ifx \showURL      \undefined \def \showURL       {\relax}        \fi
\providecommand\bibfield[2]{#2}
\providecommand\bibinfo[2]{#2}
\providecommand\natexlab[1]{#1}
\providecommand\showeprint[2][]{arXiv:#2}

\bibitem[Bojanowski et~al\mbox{.}(2016)]%
        {fasttext}
\bibfield{author}{\bibinfo{person}{Piotr Bojanowski}, \bibinfo{person}{Edouard
  Grave}, \bibinfo{person}{Armand Joulin}, {and} \bibinfo{person}{Tomas
  Mikolov}.} \bibinfo{year}{2016}\natexlab{}.
\newblock \bibinfo{title}{Enriching Word Vectors with Subword Information}.
\newblock
\newblock
\urldef\tempurl%
\url{https://doi.org/10.48550/ARXIV.1607.04606}
\showDOI{\tempurl}


\bibitem[Cui and Shrouty(2020)]%
        {pin2interest}
\bibfield{author}{\bibinfo{person}{Song Cui} {and} \bibinfo{person}{Dhananjay
  Shrouty}.} \bibinfo{year}{2020}\natexlab{}.
\newblock \showarticletitle{Interest Taxonomy: A knowledge graph management
  system for content understanding at Pinterest}.
\newblock  (\bibinfo{year}{2020}).
\newblock
\urldef\tempurl%
\url{https://medium.com/pinterest-engineering/interest-taxonomy-a-knowledge-graph-management-system-for-content-understanding-at-pinterest-a6ae75c203fd}
\showURL{%
\tempurl}


\bibitem[Devlin et~al\mbox{.}(2018)]%
        {bert}
\bibfield{author}{\bibinfo{person}{Jacob Devlin}, \bibinfo{person}{Ming-Wei
  Chang}, \bibinfo{person}{Kenton Lee}, {and} \bibinfo{person}{Kristina
  Toutanova}.} \bibinfo{year}{2018}\natexlab{}.
\newblock \bibinfo{title}{BERT: Pre-training of Deep Bidirectional Transformers
  for Language Understanding}.
\newblock
\newblock
\urldef\tempurl%
\url{https://doi.org/10.48550/ARXIV.1810.04805}
\showDOI{\tempurl}


\bibitem[Kanagal et~al\mbox{.}(2012)]%
        {yahoorecommender}
\bibfield{author}{\bibinfo{person}{Bhargav Kanagal}, \bibinfo{person}{Amr
  Ahmed}, \bibinfo{person}{Sandeep Pandey}, \bibinfo{person}{Vanja Josifovski},
  \bibinfo{person}{Jeff Yuan}, {and} \bibinfo{person}{Lluis Garcia-Pueyo}.}
  \bibinfo{year}{2012}\natexlab{}.
\newblock \showarticletitle{Supercharging Recommender Systems using Taxonomies
  for Learning User Purchase Behavior}.
\newblock  (\bibinfo{year}{2012}).
\newblock
\urldef\tempurl%
\url{https://doi.org/10.48550/ARXIV.1207.0136}
\showDOI{\tempurl}


\bibitem[Maheswari(2019)]%
        {zillowtrulia}
\bibfield{author}{\bibinfo{person}{Jyoti~Prakash Maheswari}.}
  \bibinfo{year}{2019}\natexlab{}.
\newblock \showarticletitle{My Internship at Zillow Group AI Part 1: Attribute
  Recognition in Real Estate Listings}.
\newblock  (\bibinfo{year}{2019}).
\newblock
\urldef\tempurl%
\url{https://medium.com/zillow-tech-hub/my-internship-at-zillow-group-ai-part-1-attribute-recognition-in-real-estate-listings-7d2f92009552}
\showURL{%
\tempurl}


\end{thebibliography}

\section*{Speaker Bios}
\label{appendix:speakerbio}
\textbf{Anish Khazane} is an Applied Scientist on the Home Understanding AI team at Zillow Group. He primarily focuses on building scalable machine learning models that enable rich content understanding and home recommendations for millions of users. Prior to Zillow, he worked at Capital One on building deep neural networks for representing financial transactions for merchant recommendations and advanced language models for the company's Eno chatbot. He holds a M.S in Computer Science from the Georgia Institute of Technology and a B.S in Computer Science from the University of California, Berkeley.
\\
\\
\textbf{Zachary Harrison} is an Applied Scientist at Zillow Group where he works on the Home Understanding AI team. His work focuses on developing machine learning models for content extraction and recommendations for Zillow’s customers. He holds a M.S. in Computer Science from the University of Massachusetts Amherst and a B.S in Computer Science and Computer Engineering from the University of Wisconsin-Madison. 

\newpage
\appendix
\section{Appendix: Additional Evaluation of Taxonomy Quality}

\subsection{Evaluation of Subtree Similarity}
\begin{table}[h]
  \caption{Embedding Based Subtree Similarity (add more detailed caption)}
  \label{tab:freq}
  \begin{tabular}{ccccc}
    \toprule
    
    \multicolumn{1}{p{2cm}}{\centering High Level Tree} & \multicolumn{1}{p{2cm}}{\centering High Level Tree Score} & \multicolumn{1}{p{3cm}}{\centering Subtree} & \multicolumn{1}{p{2cm}}{\centering Subtree Score} & \multicolumn{1}{p{2cm}}{\centering Subtree Size}\\
    \midrule
    \multicolumn{1}{p{2cm}}{\centering Random} & \multicolumn{1}{p{2cm}}{\centering 0.545} & \multicolumn{1}{p{3cm}}{\centering -} & \multicolumn{1}{p{2cm}}{\centering -} & \multicolumn{1}{p{2cm}}{\centering 62} \\
    \midrule
    \multicolumn{1}{p{2cm}}{\centering Exterior} & \multicolumn{1}{p{2cm}}{\centering 0.658} & \multicolumn{1}{p{3cm}}{\centering Swimming Pool} & \multicolumn{1}{p{2cm}}{\centering 0.778} & \multicolumn{1}{p{2cm}}{\centering 576} \\
    \midrule
    \multicolumn{1}{p{2cm}}{\centering Interior} & \multicolumn{1}{p{2cm}}{\centering 0.633} & \multicolumn{1}{p{3cm}}{\centering Natural Light} & \multicolumn{1}{p{2cm}}{\centering 0.755} & \multicolumn{1}{p{2cm}}{\centering 338} \\
    \midrule
    \multicolumn{1}{p{2cm}}{\centering Location} & \multicolumn{1}{p{2cm}}{\centering 0.554} & \multicolumn{1}{p{3cm}}{\centering Dining and Drinking} & \multicolumn{1}{p{2cm}}{\centering 0.748} & \multicolumn{1}{p{2cm}}{\centering 214} \\
  \bottomrule
\end{tabular}
\label{embeddingsimilarity}
\end{table}
We can plot the embedding representations of every tree node in the taxonomy to order to look for subtree clustering. Figure 1 shows this clustering of node embeddings with cosine similarity as the distance measure. This visualization technique helps with pruning the taxonomy in case there are needed hierarchy modifications, such as merging similar trees or breaking up larger ones. The location subtree, for example, is quite a large tree covering many different keywords. Furthermore, Figure 2 displays an even clearer clustering of location topics from the next level of the location subtree.
We can now define a subtree similarity metric with the following protocol (i) take all unique node pairs within each subtree (ii) calculate cosine similarity between their respective embedding representations and (iii) average these similarities across all pairs in a subtree. The average similarity represents the connectivity of a subtree, with larger values indicating a more topic specific subtree.

In Table \ref{embeddingsimilarity}, we display the results from computing this metric over different depths of subtrees (e.g exterior to swimming pool), with a comparison against a baseline approach of randomly creating edges between nodes and parents in the taxonomy. As seen in the table, the average similarity at all depths of different subtrees beats the random approach. We also observe that the further we move down a subtree, the higher the similarity score is which means more specific or semantically similar neighboring nodes. 

We also note that the Location subtree has quite a low score, barely beating the randomly generated tree. This is due to the large size of this tree $(4081)$ as well as the diverse topics this tree covers. For example, attractions in a city such as restaurants, museums, or sports stadiums will be quite different from attractions in a more rural setting which may include mountains, hiking trails or lakes. 
\begin{figure}[h]
 \centering
  \begin{subcaption}
    \centering 
    \parbox[b]{\textwidth}{
    \centering
    \includegraphics[width=180px]{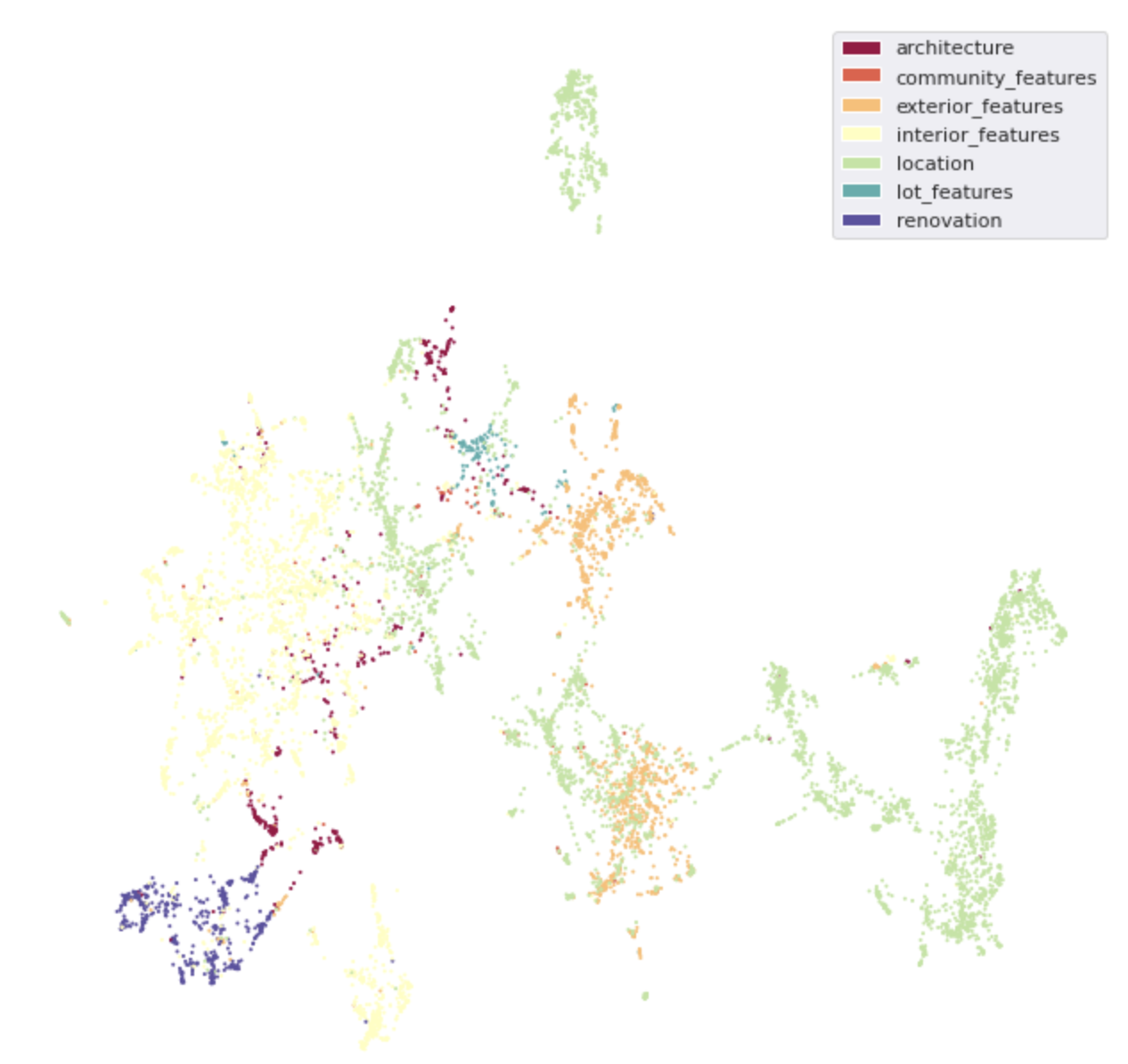}
    \caption{Visualization of tag embeddings labeled based on their highest level subtree.}}
    \parbox[b]{\textwidth}{
    \centering
    \includegraphics[width=160px]{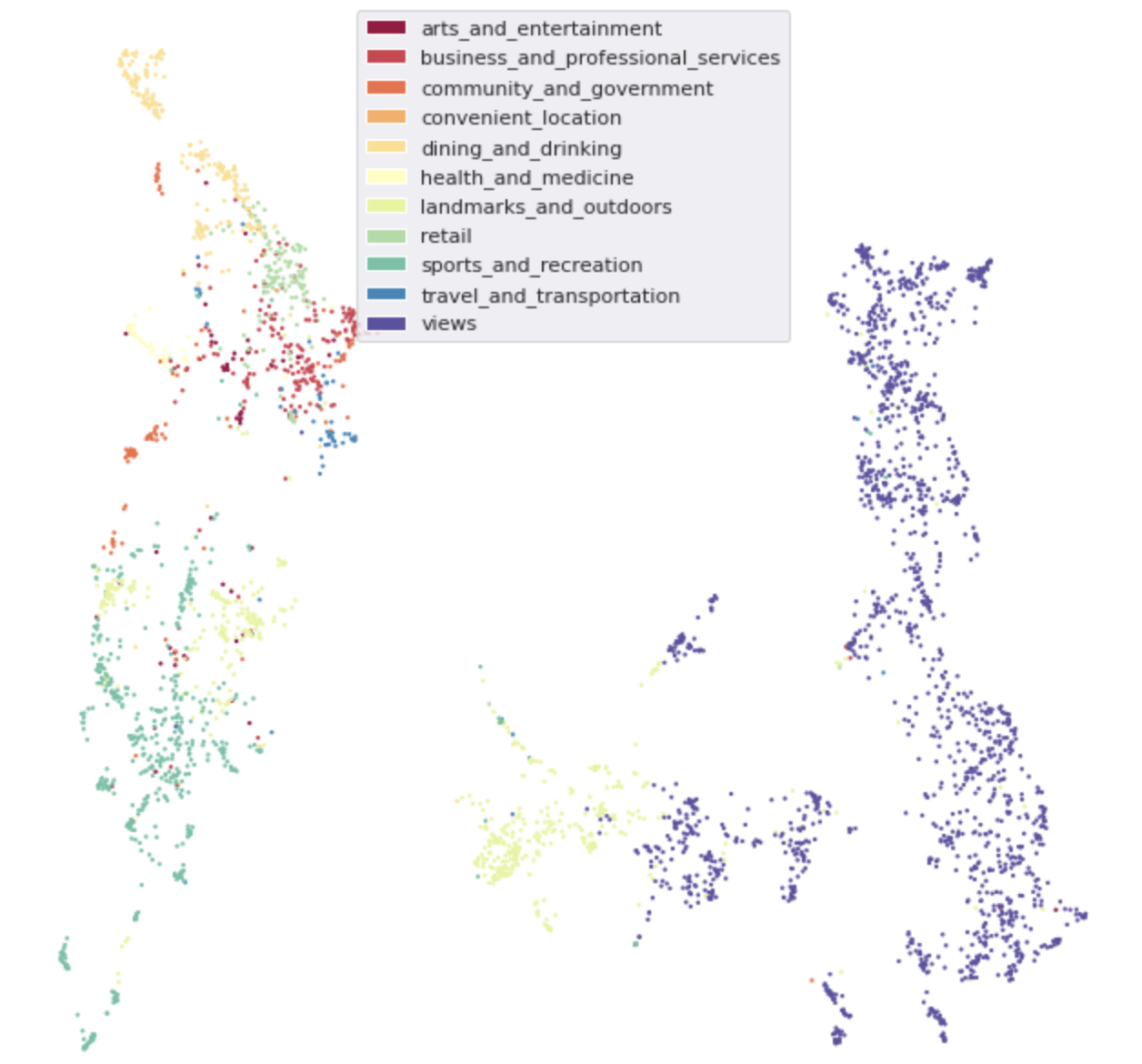}
    \caption{Visualization of Location tag embeddings labeled based on subtrees within Location.}}
  \end{subcaption}
\end{figure}

\subsection{Visual Representation of Taxonomy-based Recommendations}

\begin{figure}[h]
  \centering
  \includegraphics[width=500px]{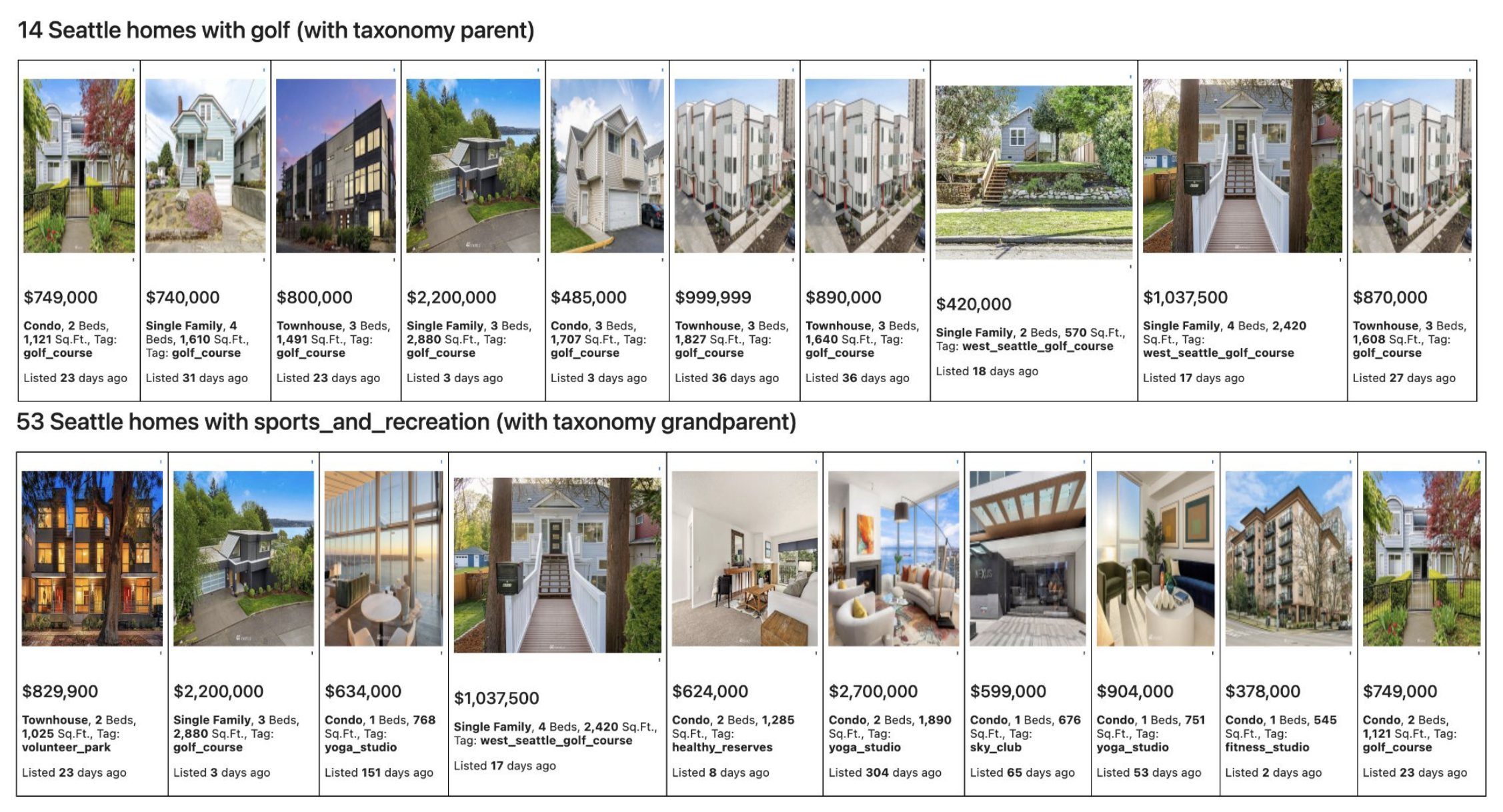}
  \caption{Visual example of recommendations at the parent and grandparent level of taxonomy for the search query, "golf". As seen in the image, categorizing the query under different levels of resolutions allows for a larger and more diverse number of recommendations. }
  \Description{A woman and a girl in white dresses sit in an open car.}
\end{figure}


\end{document}